\documentclass[sigconf, nonacm]{acmart}
\usepackage{float}
\usepackage{caption}
\newcommand\vldbdoi{XX.XX/XXX.XX}

\newcommand\vldbavailabilityurl{URL_TO_YOUR_ARTIFACTS}
\newcommand\vldbpagestyle{plain}
\begin{document}
%\title{From deep learned model with synthetic data to V$_2$O$_5$ nanowires segmentation}

\title{A deep learned nanowire segmentation model using synthetic data augmentation}

%%
%% The "author" command and its associated commands are used to define the authors and their affiliations.
\author{B. Lin*, N. Emami*, D. A. Santos**, Y. Luo**, S. Banerjee**, B.-X. Xu*}

\affiliation{%
  \institution{*Institute of Materials Science, Technische Universit\"at Darmstadt}
  \streetaddress{Otto-Berndt-Str. 3}
  \city{64287 Darmstadt }
  \country{Germany}
}
\email{b.lin@mfm.tu-darmstadt.de}

\affiliation{%
  \institution{**Department of Chemistry, Texas A\&M University,}
  \streetaddress{College Station}
  \city{TX 77843-3255}
  \country{USA}
}
\email{banerjee@chem.tamu.edu}

%%
%% The abstract is a short summary of the work to be presented in the
%% article.
\begin{abstract}
Automatized object identification and feature analysis of experimental image data are indispensable for data-driven material science; deep  learning-based segmentation algorithms have been shown to be a promising technique to achieve this goal. However, acquiring of high-resolution experimental images and assigning labels in order to  train such algorithms is challenging and costly in terms of both time and labor expense. In the present work, we apply synthetic images, which resemble the experimental image data in terms of geometrical and visual features, to train  state-of-art deep learning-based Mask R-CNN algorithms to segment vanadium pentoxide (V$_2$O$_5$) nanowires, a canonical cathode material within optical intensity-based images from spectromicroscopy. The performance evaluation demonstrates that even though the deep learning model is trained on pure synthetically generated structures, it can segment real optical intensity-based spectromicroscopy images of complex V$_2$O$_5$ nanowire structures in overlapped particle networks, thus providing reliable statistical information. The model can further be used to segment nanowires in scanning electron microscopy (SEM) images, which are fundamentally different from the training dataset known to the model. The proposed methodology of using a purely synthetic dataset to train the deep learning model can be extended to any optical intensity-based images of variable particle morphology, extent of agglomeration, material  class, and beyond.

\end{abstract}

\maketitle

%%% do not modify the following VLDB block %%
%%% VLDB block start %%%
\pagestyle{\vldbpagestyle}
%%%\begingroup\small\noindent\raggedright\textbf{PVLDB Reference Format:}\\
%\vldbauthors. \vldbtitle. PVLDB, \vldbvolume(\vldbissue): \vldbpages, \vldbyear.\\
%\href{https://doi.org/\vldbdoi}{doi:\vldbdoi}
%\endgroup
% \begingroup
% \renewcommand\thefootnote{}\footnote{\noindent
% This work is licensed under the Creative Commons BY-NC-ND 4.0 International License. Visit \url{https://creativecommons.org/licenses/by-nc-nd/4.0/} to view a copy of this license.
% \href{https://doi.org/\vldbdoi}{doi:\vldbdoi} \\
% }\addtocounter{footnote}{-1}\endgroup
%%% VLDB block end %%%

%%% do not modify the following VLDB block %%
%%% VLDB block start %%%
%\ifdefempty{\vldbavailabilityurl}{}{
%\vspace{.3cm}
%%\begingroup\small\noindent\raggedright\textbf{Data and Code Availability:}\\
%The source code, data, and/or other artifacts have been made available at \url{\vldbavailabilityurl}.
%\endgroup
%}
%%% VLDB block end %%%

\section{Introduction}
Understanding the design rules that dictate materials chemistry is critical to enabling the rational design of energy storage systems.  Moreover, connecting single-entity and ensemble measurements is paramount to understanding how structure-function relationships propagate across length scales and dictate the performance of hierarchical systems in battery materials \cite{baker2018perspective,li2020peering}. The ability to probe a multitude of contrast mechanisms from a single measurement has enabled many insights into the working principles of electrochemically active materials \cite{wolf2017visualization}. Spectromicroscopy techniques such as scanning transmission X-ray microscopy (STXM) and X-ray ptychography, for example, leverage X-absorption and scattering events to capture morphological and electronic structure information which can be colocalized at the nanometer level to provide chemical maps for a region of interest\cite{yu2018three,santos2020bending,andrews2020curvature}. The application of such information-rich measurements to particle networks has been limited, in part, due to the complexity of extracting morphological and chemical features from large and complex datasets.Dataset dimensionality reduction techniques such as principal component analysis considerably improve the ease of deciphering chemical markers often contained within spectra~\cite{lerotic2014mantis}. Nevertheless, there is a need for more efficient and effective workflows to obtain size and shape descriptors that can be utilized with chemical information to explore physio-chemical phenomena as a function of various descriptors.

In recent years, image segmentation algorithms that leverage the parallel processing capability of neural networks have garnered significant attention because of their potential to enable automated image analysis ~\cite{lecun2015deep,he2016deep}. For example, the well-received Mask R-CNN algorithm~\cite{he2017mask} is now utilized routinely for segmentation tasks. Common Object in Context (COCO)~\cite{lin2014microsoft} and PASCAL Visual Object Classes (VOC)~\cite{pascal-voc-2010} have been developed in concert to train and benchmark the performance of algorithms for object detection, semantic segmentation, and general classification tasks in the field of computer vision. However, the requirement of large datasets to train deep-learning algorithms has been challenging to meet with experimental microscopy data in the material-chemistry community due to an inherent complexity in generation and the time-consuming nature of human annotation. Nevertheless, deep-learning algorithms based on empirical data and human annotation have been developed for several material classes, such as graphene flakes ~\cite{masubuchi2020deep} imaged by optical microscopy, carbon nano-fibers~\cite{frei2021fiber} imaged by SEM and a further collection of electron microscopy images for various material class ~\cite{yildirim2021bayesian}. Similarly, attempts to augment real image datasets of polycrystalline grains~\cite{ma2020data}, or usage of image rendering techniques on nano-particles images~\cite{decost2017characterizing, mill2020synthetic} to counteract the prohibitive data acquisition step, have been proven to be successful.

To overcome the challenges associated with limited training data we have developed a deep learning model, based on the Mask R-CNN algorithm, which has been trained entirely on synthetically generated microstructures. By  applying an optical density compression step, the algorithm can segment and obtain statistical information from both electron and X-ray microscopy images of vanadium pentoxide
(V$_2$O$_5$), a canonical cathode material. From an image segmentation perspective, the V$_2$O$_5$ nanoparticle dispersions shown in Fig.\ref{fig:image_data} represent a fundamentally case study against "ideal" monodispersion of nanoparticles characterized by a nearly spherical geometry for which the automated size determination process is documented~\cite{ruhle2021workflow, yildirim2021bayesian}. Through the lens of chemistry-mechanics coupling, V$_2$O$_5$ appears as a fascinating case study of image analysis, specifically, it is well known that the patterns of lithiation in these systems are strongly modified by dimensional and morphological features such as particle geometry, curvature, and interconnects~\cite{santos2020bending,andrews2020curvature,horrocks2013finite}.

In the present work, we consider three different types of microscopy images generated by X-ray ptychography, scanning transmission X-ray microscopy (STXM) and scanning electron microscopy (SEM) techniques. While each imaging technique is fundamentally different, the algorithm developed in this work demonstrates a remarkable robustness when segmenting nanorod-like structures. The present work is organized as follows: in section 2.1 we briefly outline the experimental methodology and introduce the experimental datasets, which are later utilized to validate the performance of the model. For the model training, we generate a series of synthetic images. The workflow of data generation is explained in section 2.2, whereas the details on the training of Mask R-CNN model are presented in section 2.3. In section 3, results on training, evaluation, and segmentation of considered microscopy images from the above mentioned imaging techniques are provided  and discussed.

%  we propose the use of MaskR-CNN model to perform the segmentation task of V$_2$O$_5$ battery nanowires as a feasibility study. But with the novelty, train the model purely on synthetically generated images. The other strength from our generated dataset are to be the individual labels that are provided to the training process, such that model can segment particles on instance level, instead of semantic level and detect individual particles. The aim of the work is firstly to check the feasibility and ability of the deep learning model, which are purely trained on synthetic dataset, for the segmentation process and further use this model to analyse the statically distributed features after particles segmentation as a post-processing step. It is shown in the work, that the initially trained model on optical-density based images are even capable of segmenting SEM images, and this shows big potentials also for other material system.

\section{Material and Methods}
\subsection{Synthesis and Imaging of Nanowires Particles }

\begin{figure}[h]
  \centering
  \includegraphics[width=0.4\textwidth]{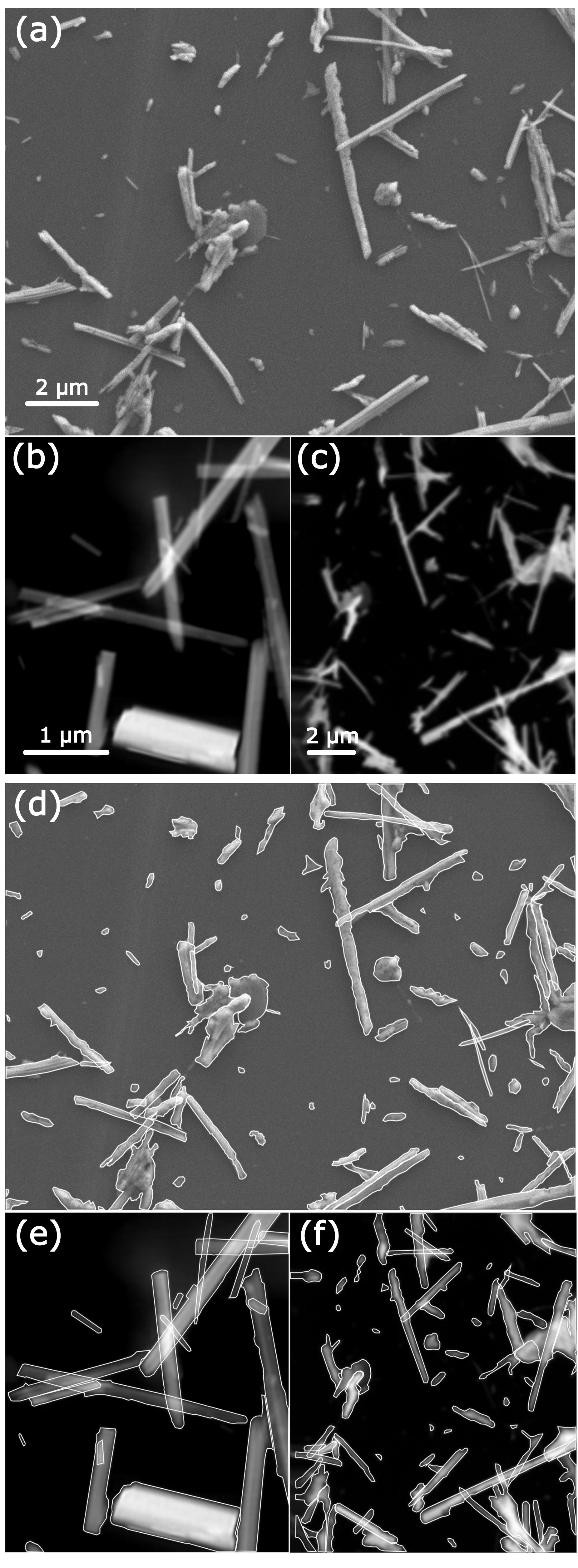}
  \caption{Overview of the microscopy images: (a) SEM (b) X-ray ptychography (c) STXM (d)-(f) Corresponding manual annotation}
  \label{fig:image_data}
\end{figure}

\subsubsection{Synthesis of V$_2$O$_5$ Nanowires}
Nanowires of $\alpha$-V$_2$O$_5$, the thermodynamic sink for binary vanadium oxides, were synthesized by a hydrothermal growth process. Briefly, V$_3$O$_7$·H$_2$O nanowires were initially prepared and calcined in air to obtain $\alpha$-V$_2$O$_5$ nanowires crystallized in the orthorhombic phase, as reported previously~\cite{santos2020bending}. Typical dimensions span from 50 to 400 nm in width and up to several microns in length. Chemical lithiation was achieved via submersion into a 0.01M n-butyllithium solution in heptane. Metastable $\zeta$-V$_2$O$_5$ nanowires were prepared by a series of hydrothermal reactions as described in previous work~\cite{andrews2018reversible}. Briefly, bulk V$_2$O$_5$ and silver acetate were hydrothermally reacted to form an intermediate $\beta$-Ag$_{0.33}$V$_2$O$_5$ product. To create the tunnel-structured $\zeta$-V$_2$O$_5$, $\beta$-Ag$_{0.33}$V$_2$O$_5$ was hydrothermally reacted with HCl in aqueous conditions to leach the Ag from the structure. For electrochemical sodiation, CR2032 coin cells were prepared under an inert argon environment. The working electrode was prepared by casting a mixture of the active material ($\zeta$-V$_2$O$_5$, 70 wt.\%), conductive carbon (Super C45, 20 wt.\%), and binder [poly(vinylidene fluoride) 10 wt.\%] dispersed in N-methyl-2-pyrrolidone onto an Al foil substrate. Sodium metal and glass fiber were used for the counter electrode and separator, respectively. For the electrolyte, 1M NaPF6 solution was prepared using a solvent mixture of ethylene carbonate and diethyl carbonate (1:1 volumetric ratio). The extent of sodiation was controlled by galvanostatic discharging using a LANHE (CT2001A) battery testing system. Cells were disassembled, washed with dimethyl carbonate (DME), and dried for 24 h in an inert argon environment.
\subsubsection{Scanning Electron Microscopy (SEM)}
Before imaging, dispersions of $\alpha$-V$_2$O$_5$ and $\zeta$-V$_2$O$_5$ nanowires were created by drop-casting onto a silicon nitride substrate~\cite{santos2020bending}. The SEM image shown in Fig.~\ref{fig:image_data}(a) was collected on a Tescan LYRA-3 instrument equipped with a Schottky field-emission source and a low aberration conical objective lens.
\subsubsection{X-ray ptychography}
X-ray ptychography measurements were performed at the coherent scattering and microscopy beamline of the Advanced Light Source in Berkeley, CA. An optic with a 60 nm outer zone width, and a 40 nm step-size of the field of view was utilized. The image shown in Fig.~\ref{fig:image_data}(b) depicts the ratio between the t$_2g$ (527 eV) and e$_g*$ (529.8 eV) fine structure features at the O K-edge which correspond to transitions from O 1s core states to O 2p states hybridized with V 3d states and is indicative of the extent of intercalation.
\subsubsection{Scanning transmission X-ray microscopy (STXM)}
The STXM measurements were performed at the spectromicroscopy beamline 10D-1 of the Canadian Light Source in Saskatoon, SK utilizing a 7 mm generalized Apple II elliptically polarizing undulator source (EPU). Here, a focused beam spot was raster-scanned across the field of view with a 35nm step size (thus determining the spatial resolution). A series of images were collected from 508 eV to 560 eV in 0.2 eV increments. The STXM image shown in Fig.~\ref{fig:image_data}(c) depicts the average absorption (optical density) contrast from 508 eV to 560 eV \cite{santos2020bending}.

\subsubsection{Human annotation of the V$_2$O$_5$ dataset}
Manual annotation of the V$_2$O$_5$ datasets was facilitated by the  web-based annotation tool Makesense.ai \cite{makesense} in a polygon format, where points along a particle border are set to form the shape of the particles (see Fig.\ref{fig:image_data}). This step is performed for every particle in the present images, and the annotated file is saved in JSON formats to serve the validation purpose. It is worth noting that some limitations of the manual annotation process such as a sensitivity to human error and a dependence on spatial resolution of the native images naturally exist. Further sources of error stem from the inherent complexity of the dispersion, which results in many instances where particles are overlapped.

\subsection{Synthetic dataset generation}
\begin{figure}[H]
  \centering
  \includegraphics[width=1.0\columnwidth]{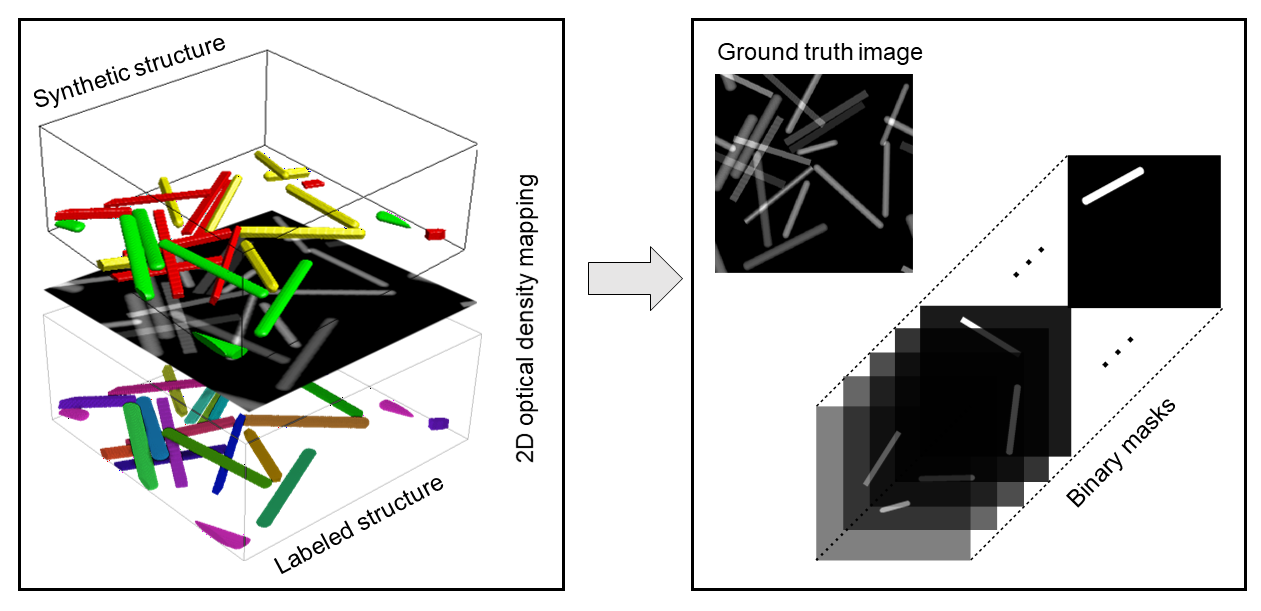}
  \caption{Synthetically generated dataset for training procedure. The 3D-microstructure was compressed to create optical density-based image as input data. The individually labelled nanowires in correlation with the optical density-based image are then used to create the binary masks for output data.}
  \label{fig:sythetic_data}
\end{figure}
To generate synthetic image datasets reminiscent of the V$_2$O$_5$ experimental datsets  particles, we have developed a random nanowire generator using the software Geodict\textsuperscript\textregistered. In the generation step, for each training sample, the number of particles, length, shape distribution was specified to create 3D voxel-based structures, Fig.~\ref{fig:sythetic_data}. The chosen size of the domain was 512x512x200 (WxBxH). For the present work, the number of particles, diversity of morphology, and resolution, approximate the experimental information contained in the X-ray ptychography data in Fig~\ref{fig:image_data}(b). Higher resolution i.e. larger domain size, can be chosen at cost of longer generation time and image file size.  The height of the domain for the synthetic 3D microstructure was chosen such that it exceeded the total height of the overlapped nanowires.  Further, the nanowires were internally enumerated and deposited one after another. This workflow ensures that the labels are predetermined, thus bypassing the need for human annotation at a later phase. In order to mirror the transmission intensity generated by X-ray ptychography, an optical density compression step was applied to emulate thickness information. Here, voxels were first compressed along the out-of-plane direction then summed and divided by the total thickness of the microstructure. The pixel values comprising the projected optical density image map in the in-plane directions were therefore normalized, and further transformed to a gray color scale, expressed as a value from 0 to 255. Accordingly, regions where two or more particles are overlapped can be distinguished by a sudden change in optical density (i.e. pixel intensity). %\textcolor{red}{Here, we treated the optical intensity from X-ray ptychography image, which corresponds to the state of intercalation simply as optical density as in the STXM image and assume that the state of intercalation and average absorption are interchangeable for the deep learning algorithm from the visual segmentation perspective.}
In a subsequent step, a standard Gaussian filter (filter size of 2) was applied to the images to account for blurring of the particle edges. As shown in Fig.~\ref{fig:sythetic_data}, the ground truth sample was split into individual binary masks for each nanowire contained in the synthetic ground truth image. Therefore, for each training sample, the dataset consists of one input image and N output binary mask images with N as the number of particles in that image. These binary mask images can be further used to obtain statistical information about the morphology descriptors.  Before evaluation, initial dataset size of 250 images  were generated for training purposes. After an initial evaluation, additional images were generated in recursive steps involving a greater diversity of particle morphology to closely replicate the experimental data. A final dataset of 1000 synthetic images were obtained for the training and validation steps. Details on the training dataset can be found in the section data and code availability.

\subsection{Deep learning procedures}
\begin{figure}[H]
  \centering
  \includegraphics[width=1.0\columnwidth]{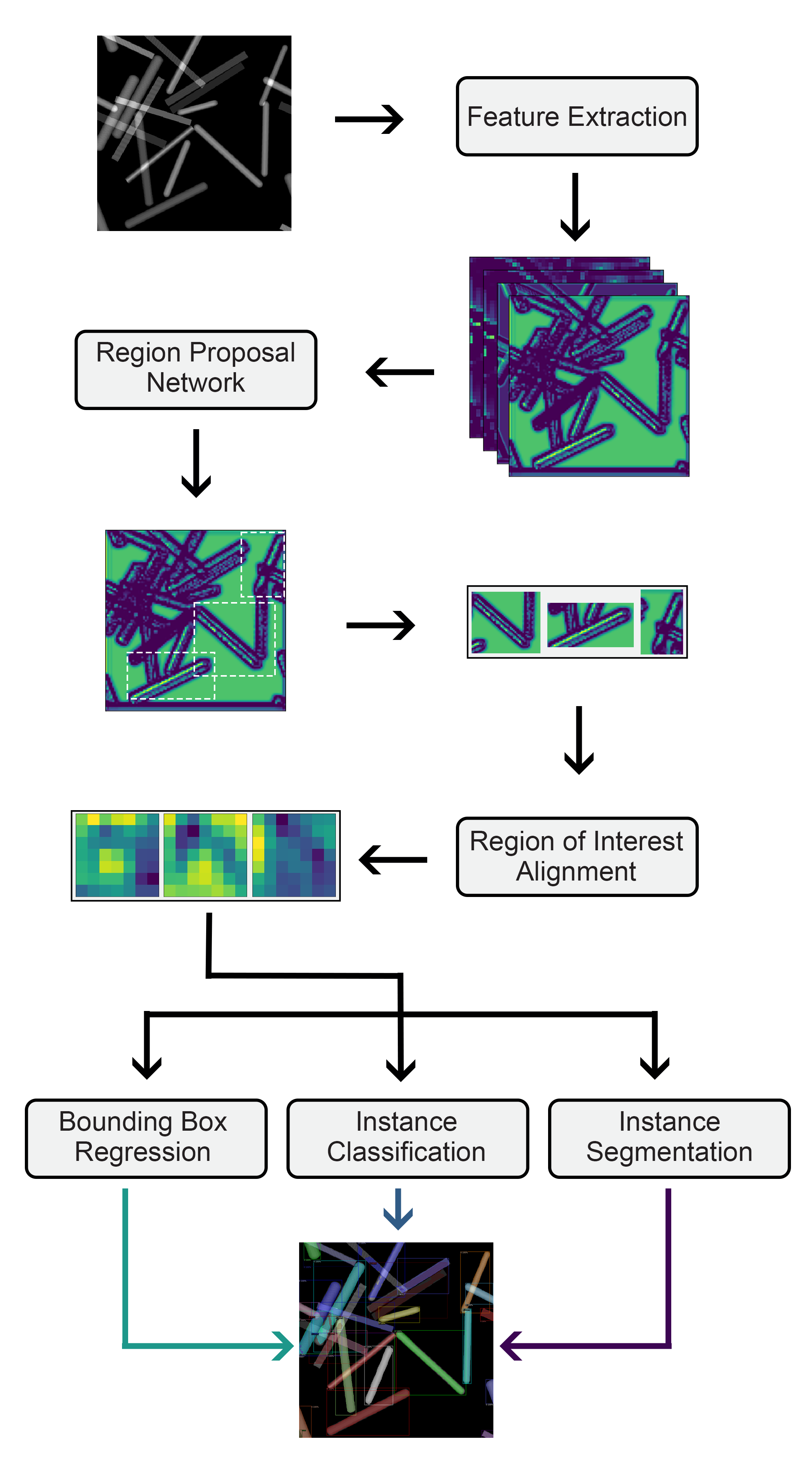}
  \caption{Workflow of the Mask-RCNN}
  \label{fig:M-RCNN-workflow}
\end{figure}
\subsubsection{Basic components}
The Mask R-CNN architecture utilized in this work is based on the Detectron2 implementation by Facebook AI Research~\cite{wu2019detectron2}. In the following section, the basic structure of Mask R-CNN model and its workflow are briefly introduced. The model architecture can be divided into 3 main parts~\cite{he2017mask}, as illustrated in Fig.~\ref{fig:M-RCNN-workflow}:

\begin{itemize}
\item Feature extraction step
\item Region of interest proposal and alignment
\item Overhead for mask and bounding box prediction and classification
\end{itemize}
%\textcolor{red}{BX: the bullet points and the numbering 1-3 are duplicated? why both are given here?}
\begin{enumerate}
\item This step is usually referred to as the backbone of the model and is constructed with multiple CNN layers. The input image is introduced and passed through the CNNs to extract representative features of the entire image. The CNN layers are usually deep and contain most of the model weights updated during the training steps. The backbone used here can easily be tailored to the desired segmentation task in order to improve speed and performance. The backbone used for the training in this work is a FPN with the ResNet-50 network pretrained on COCO-dataset \cite{he2017mask}.

\item This step in the model is designed to identify and extract instances from feature maps produced by the backbone. The Region proposal network (RPN) achieves this by generating a series of region of interest (ROIs), each encapsulating a single instance. The model generates hundreds of ROIs and an associated a confidence score to quantify the probability of encompassing an object for the given ROI. After filtering and modifying the coordinates of each ROI, RPN advances portions of the feature map (corresponding to each ROI) with a fixed size to the model prediction head in order to determine the properties (e,g, bounding box, and instance mask, etc.) of each instance.

\item The so-called prediction heads are functions that predicts the characteristics of that proposed instance. For object detection purpose, most common R-CNN structures typically provide two instance heads, namely bounding box regression head, which draws a bounding box around an instance, and the instance classification head to classify the object class. The typical prediction head of Mask R-CNN is therefore a extension of R-CNN models with the mask segmentation, in which a binary mask is generated to label the predicted instance.
\end{enumerate}

\subsubsection{Loss functions}
 During training, the difference between model prediction and ground truth should be minimized. This optimization procedure requires the definition of a function to perform this calculation,  -usually referred to a loss function or a cost function. Typically, in neural networks, the optimization goal is to minimize this loss function. Different loss functions can be used for different tasks based on the input data and the desired output of the model. In the Mask R-CNN model used in this work, the defined loss function is based on the summation of 3 individual loss functions~\cite{he2017mask}:
\begin{equation}
    \mathcal{L} = \mathcal{L}_{bbox} +\mathcal{L}_{cls} +\mathcal{L}_{mask}
\end{equation}\,
where $\mathcal{L}_{bbox}$ is smooth $L^1$ loss function used for predicting bounding box coordinates, while $\mathcal{L}_{cls}$ is a cross entropy loss function to measure the classification of multiple classes and returns a probability between 0 and 1. In the case of a binary segmentation (in this case, the nanowire and the background), the cross-entropy loss function can be written as:
\begin{equation}
    \mathcal{L}_{cls} = -(y \log{(p)}+\log{(1-y)}\log{(1-p)})
\end{equation}
where the $\log $ is the natural log, $y$ is a binary value (0 or 1) indicating the class of observation, and $p$ is the predicted probability for the given observation. As for $\mathcal{L}_{mask}$, it is a binary cross-entropy for the generated binary mask of size $m \times m$ for each ROI.

\subsubsection{Input augmentations}
In order to introduce further variability in the dataset, standard augmentation techniques have been applied as the dataset is fed forward to the dataloader. Note here, we do not create an additional dataset in the training process.  To account for intrinsic variability in the contrast and brightness of experimental microscopy datasets, we applied a random brightness and a random contrast filter ranging from 0.7 to 1.2. A random flip in both horizontal and vertical direction with a probability of 0.5 has been applied.
\subsubsection{Training}
It is important to note that while the synthetic datasets were modeled after experimental microscopy images, the training steps for the model developed in this work were performed solely on synthetic datasets. The stochastic gradient descent method was used with the default setting provided by the Detectron2 implementation of Mask R-CNN algorithm~\cite{wu2019detectron2}. The hyperparameters modified in the parameter study can be found in Tab.~\ref{prsearch}. The training was performed on 4 A100/V100 Nvidia GPUs at Lichtenberg Cluster, TU Darmstadt for a total time of 5-10 hours with a batch size of 8 images per GPU. The training time here should only provide an approximation; Study and optimization on the performance speed was not the objective of this work.
\subsubsection{Evaluation metrics}
To evaluate the segmentation results, we make use of three metrics. The segmentation accuracy defined in Eq.~\ref{Eq:segmentation accuracy} is used here as  simple pixel-wise correctness, which only examines the true predicted pixels with predictions made from all the objects in the foreground. TP denotes the true positive, FP the false positive, and FN the false negative predictions. Note that this metric is only applied to the foreground, which differs from the common use considering the background pixels likewise.

\begin{equation}
\mathrm{Accuracy} = \frac{\mathrm{TP}}{\mathrm{TP} + \mathrm{FP} + \mathrm{FN} }
\label{Eq:segmentation accuracy}
\end{equation}\,

The second evaluation scheme is according to COCO dataset~\cite{lin2014microsoft} based on mean average precision (mAP). This is introduced briefly in the following section. Firstly, to confirm a correct prediction of bounding box or mask, intersection over Union is used (IoU). It is defined by the area of intersection between bounding boxes divided by their union as shown in Fig.\ref{fig:iou}. Predictions are true positive, if IoU is higher than a given threshold, and false negative if that is lower than that threshold. The most common thresholds used are IoU$>50$ ($\mathrm{AP_{50}}$) and IoU$>75$ ($\mathrm{AP_{75}}$).

\begin{figure}[H]
  \centering
  \includegraphics[width=0.9 \columnwidth]{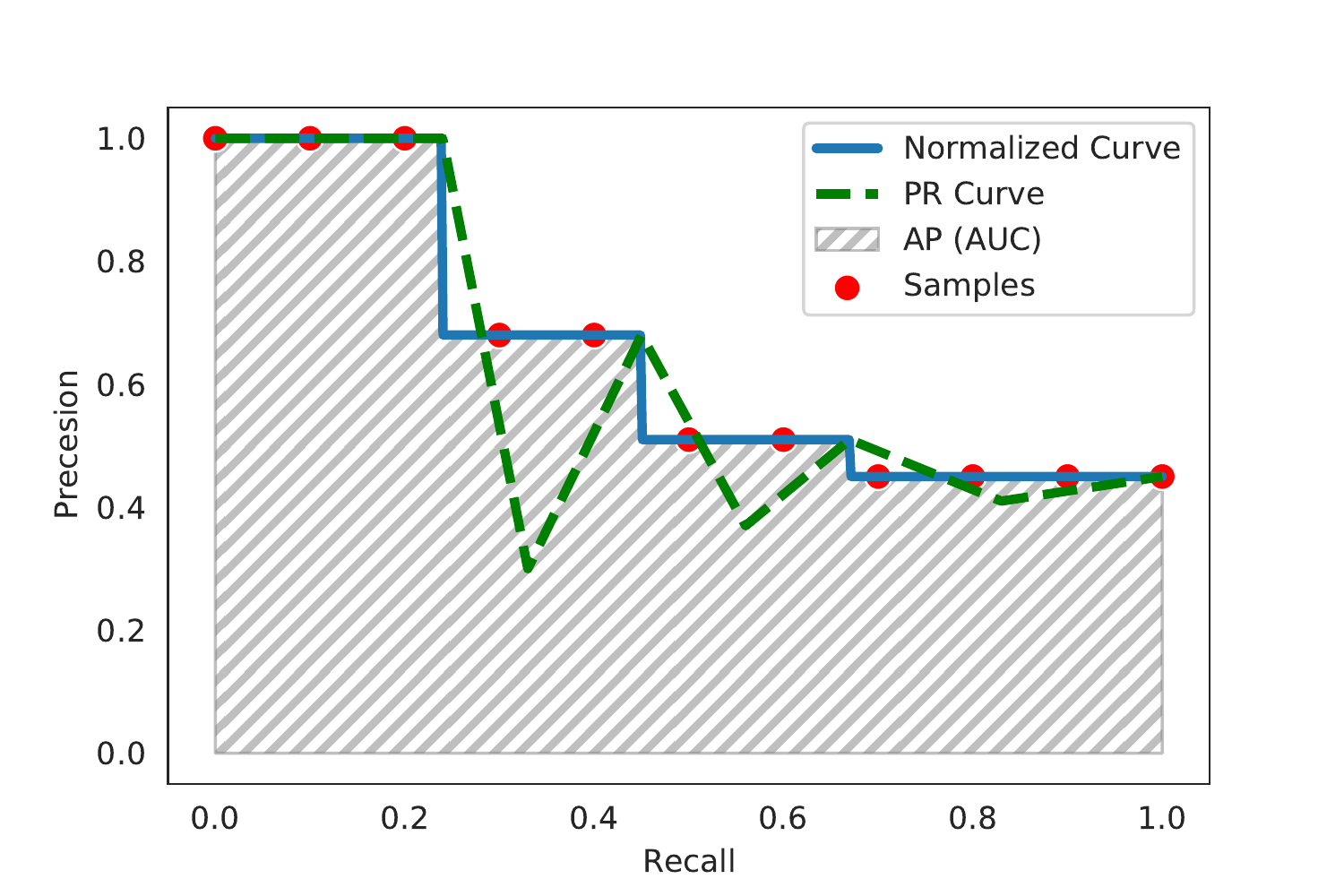}
  \caption{PR Curve}
  \label{fig:PR_curve}
\end{figure}

To further understand mAP, precision and recall are defined as follows:
\begin{figure}
  \centering
  \includegraphics[width=0.9\columnwidth]{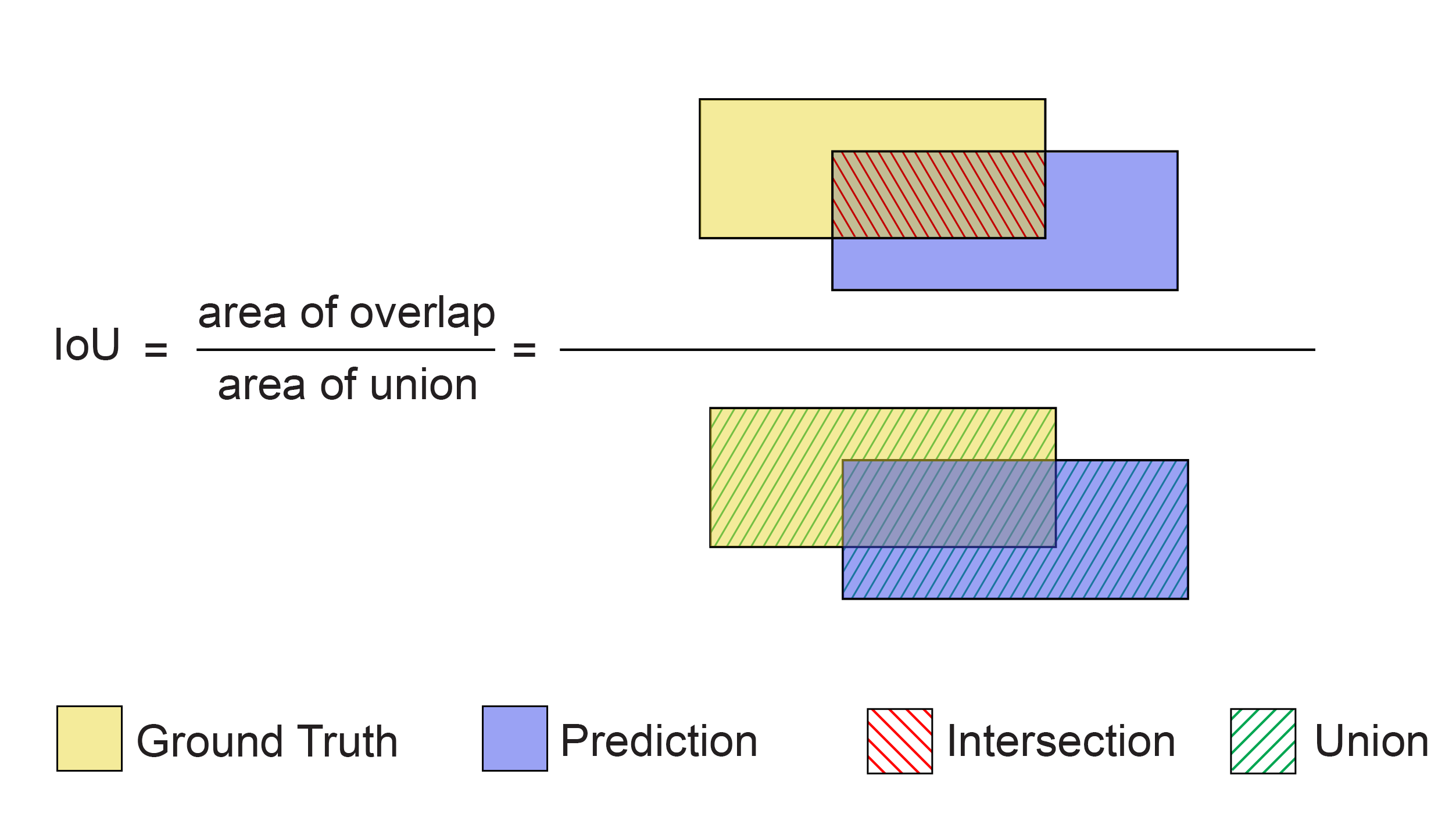}
  \caption{Intersection over union}
  \label{fig:iou}
\end{figure}
\begin{equation}
\mathrm{Precision} = \frac{\mathrm{TP}}{ \mathrm{TP} + \mathrm{FP}} \
\end{equation}\,
\begin{equation}
\mathrm{Recall} = \frac{\mathrm{TP}}{\mathrm{TP} + \mathrm{FN}}
\end{equation}\,
Recall is considered as true positive prediction rate, i.e., the ratio between true positive predictions and all ground truths. Precision is defined as the ratio between true positive predictions and all predictions that are made. Further, the obtained precision and recall are plotted to obtain the so-called precision-recall (PR) curve with the area under it referred to as average precision (AP). In VOC2010~\cite{pascal-voc-2010}, a modified PR curve was introduced, where precision for a given recall $r$ is set to the maximum precision for any $\Tilde{r} \geq r$. Afterward, the AP can be computed by numerical integration for the area under the curve (AUC) as shown in Fig.~\ref{fig:PR_curve}. mAP is later defined as the average of AP for all classes in each image, if more than one class is present.
%A model is therefore performing reasonable if recall increases with slowly decreasing precision.
%The average precision is defined as the area under the precision-recall curve ($PR $ curve). x-axis representing recall and y-axis representing precision.

For VOC, usually IoU~$ > 50$ is considered considered true positive prediction, which results in true predictions with any IoU higher than 0.5 contributing equally to the AP. To rectify this problem, COCO uses different thresholds for IoU ranging from 0.5 to 0.95 with a step size of 0.05 and then reports the average of all computed APs to varying thresholds as mAP, see Eq.\ref{eq:coco}. In this work, we use AP for $\mathrm{mAP_{coco}}$ and assume the difference is clear from the context. COCO evaluator also reports more detailed results based on the scale of the detected objects.
\begin{equation}
\mathrm{mAP_{coco}} = \frac{\mathrm{mAP_{50}}+\mathrm{mAP_{55}}+...+\mathrm{mAP_{95}}}{10}
\label{eq:coco}
\end{equation}
With reported $\mathrm{AP_{small}}$ for objects with area smaller than $32^2$ pixels, $\mathrm{AP_{medium}}$ for objects with area between $32^2$ and $69^2$ pixels and $\mathrm{AP_{large}}$ for objects with area greater than $69^2$ pixels, one can evaluate the model performance on segmenting objects in different scales.

%one thing to note is that COCO uses $[0:0.01:1]$ for recall when computing AUC in the PR curve, which means to take the maximum precision value to their right for 101 recall values in the given range rather than 11 points in VOC.

Finally, given the fundamental motivation to extract particle statistics from image datasets, the performance of the model is further evaluated based on the accuracy of the predicted statistics.. Computation is made based on the segmented masks of each particle. Statistical information obtained from the predictions are then compared to the manually annotated results.

\begin{table*}[ht]
  \caption{Hyperparameter study in the training procedure}
  \label{tab:coco_training}
  \begin{tabular}{ccccccccc}
    \toprule
     No. & No. Epochs & Dataset Size & Learning Rate & ROI Head & IOU THR & NMS &AP BBOX &  AP SEGM \\
    \midrule
    1 & 250 & 250 & 0.02 & 256 & 0.6 & 0.7 & 89.442 & 87.392\\
    2 & 250 & 500 & 0.02 & 256 & 0.6 & 0.7 & 91.843 & 88.878\\
    3 & 250 & 750 & 0.02 & 256 & 0.6 & 0.7 & 92.219 & 89.568\\
    4 & 250 & 1000 & 0.02 & 256 & 0.6 & 0.7 & 93.541 & 90.195\\
    5 & 500 & 250 & 0.02 & 256 & 0.6 & 0.7 & 90.949 & 87.908\\
    6 & 500 & 500 & 0.02 & 256 & 0.6 & 0.7 & 92.831 & 89.872\\
    7 & 500 & 750 & 0.02 & 256 & 0.6 & 0.7 & 93.336 & 90.034\\
    8 & 500 & 1000 & 0.02 & 256 & 0.6 & 0.7 & 93.878 & 90.254\\
    9 & 750 & 250 & 0.02 & 256 & 0.6 & 0.7 & 92.219 & 89.568\\
    10 & 750 & 500 & 0.02 & 256 & 0.6 & 0.7 & 92.820 & 90.095\\
    11 & 750 & 750 & 0.02 & 256 & 0.6 & 0.7 & 93.454 & 89.958\\
    12 & 750 & 1000 & 0.02 & 256 & 0.6 & 0.7 & 94.002 & 90.487\\
    13 & 750 & 500 & 0.01 & 256 & 0.6 & 0.7 & 93.564 & 89.72\\
    14 & 750 & 500 & 0.03 & 256 & 0.6 & 0.7 & 93.279 & 89.87\\
    15 & 750 & 500 & 0.02 & 128 & 0.6 & 0.7 & 92.582 & 89.614\\
    16 & 750 & 500 & 0.02 & 512 & 0.6 & 0.7 & 93.513 & 89.932\\
    17 & 750 & 500 & 0.02 & 256 & 0.7 & 0.7 & 93.278 & 89.945\\
    18 & 750 & 500 & 0.02 & 256 & 0.8 & 0.7 & 93.523 & 90.139\\
    19 & 750 & 500 & 0.02 & 256 & 0.6 & 0.6 & 92.902 & 89.952\\
    20 & 750 & 500 & 0.02 & 256 & 0.6 & 0.8 & 93.58 & 89.74\\

    % 1 & 200& 0.01 & 256 &  87.896 & 86.098  \\
    % 2 & 400 & 0.01 & 256 & 89.005 & 86.457  \\
    % 3 & 600 & 0.01 & 256 & 89.929 & 87.453  \\
    % 4 & 800& 0.1 & 256 & xx & xx  \\
    % 5 & 400 & 0.01 & 512 & 89.721 & 87.715   \\
    % 6 & 400 & 0.02 & 256 & 89.804 & 87.941  \\
    % 7 & 400 & 0.04 & 256 & 88.861 & 87.620  \\
    % $8^*$ & 400 & 0.01 & 256 & 89.809 & 87.941  \\
    % $9^*$ & 400 & 0.005 & 512 & 91.130 & 87.066  \\
    % $10^{**}$ & 400 & 0.01 & 512 & 92.210 &	86.461  \\
    % $11^{**}$ & 800 & 0.01 & 512 & 92.891 &	86.114  \\
    % $12^{**}$ & 400 & 0.01 & 512 & 92.639 & 85.893  \\

    \bottomrule
  \end{tabular}
  \label{prsearch}
\end{table*}
\begin{figure*}[h]
  \centering
  \includegraphics[width=0.85\textwidth]{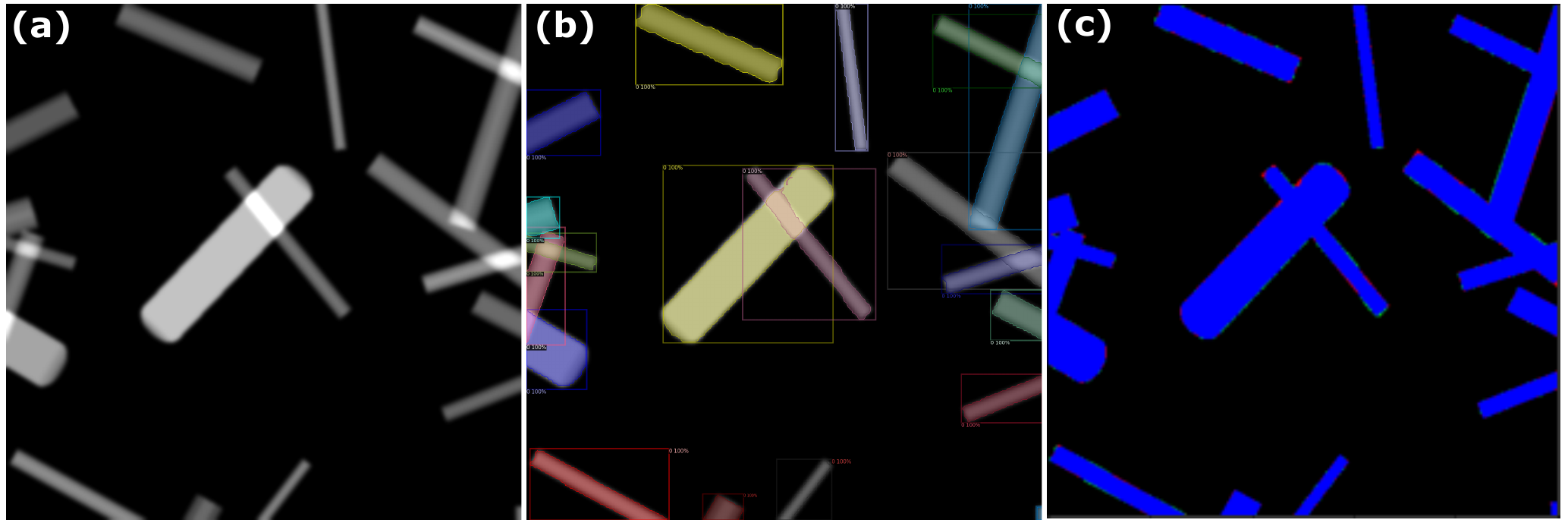}
  \caption{Model prediction on V$_2$O$_5$ nanowires within synthetic image (512x512 pixels) (a) Test image (b) Predicted instance masks with lower opacity plotted on the test images (c) Semantic binary mask, Blue: TP, Red: FN, Green: FP}
  \label{fig:Seg_syntheticData}
\end{figure*}

\section{Results and discussions}
In this section, the segmentation results are evaluated considering the aforementioned metrics. It is important to re-emphasize that the deep learning model has been trained solely on synthetic image datasets modeled after the X-ray ptychography and scanning transmission X-ray microscopy data. Consequently, the SEM image, which is distinctive in terms of contrast generation is foreign to the trained model. All images are, obtained from mentioned microscopy techniques as they are, and were not filtered for the evaluation purpose. The only pre-processing step applied to the ptychography and STXM images involves the conversion from transmission data to absorbance (optical density)~\cite{lerotic2014mantis}.  The results obtained from the deep learning model are compared to manually annotated results, which are subject to uncertainty to certain level due to visual limitation.

\begin{figure*}[h]
  \centering
  \includegraphics[width=0.9\textwidth]{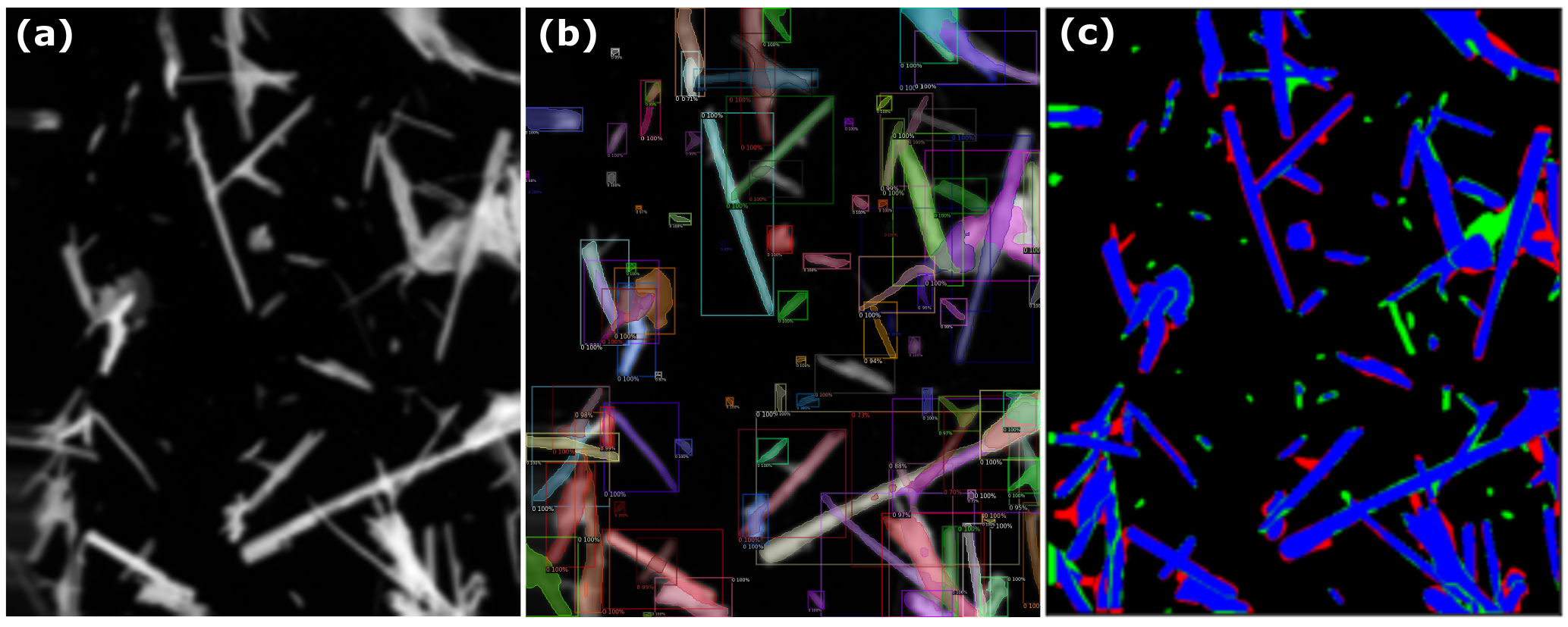}
  \caption{Model prediction on V$_2$O$_5$ nanowires images by STXM (531x449 pixels - rescaled). The best performer in AP segmentation mask Tab.~\ref{tab:score_stxm} is used to visualize the prediction masks in (b)}
  \label{fig:Seg_SXTM}
\end{figure*}

\begin{figure}[h]
  \centering
 \includegraphics[width=0.8\columnwidth]{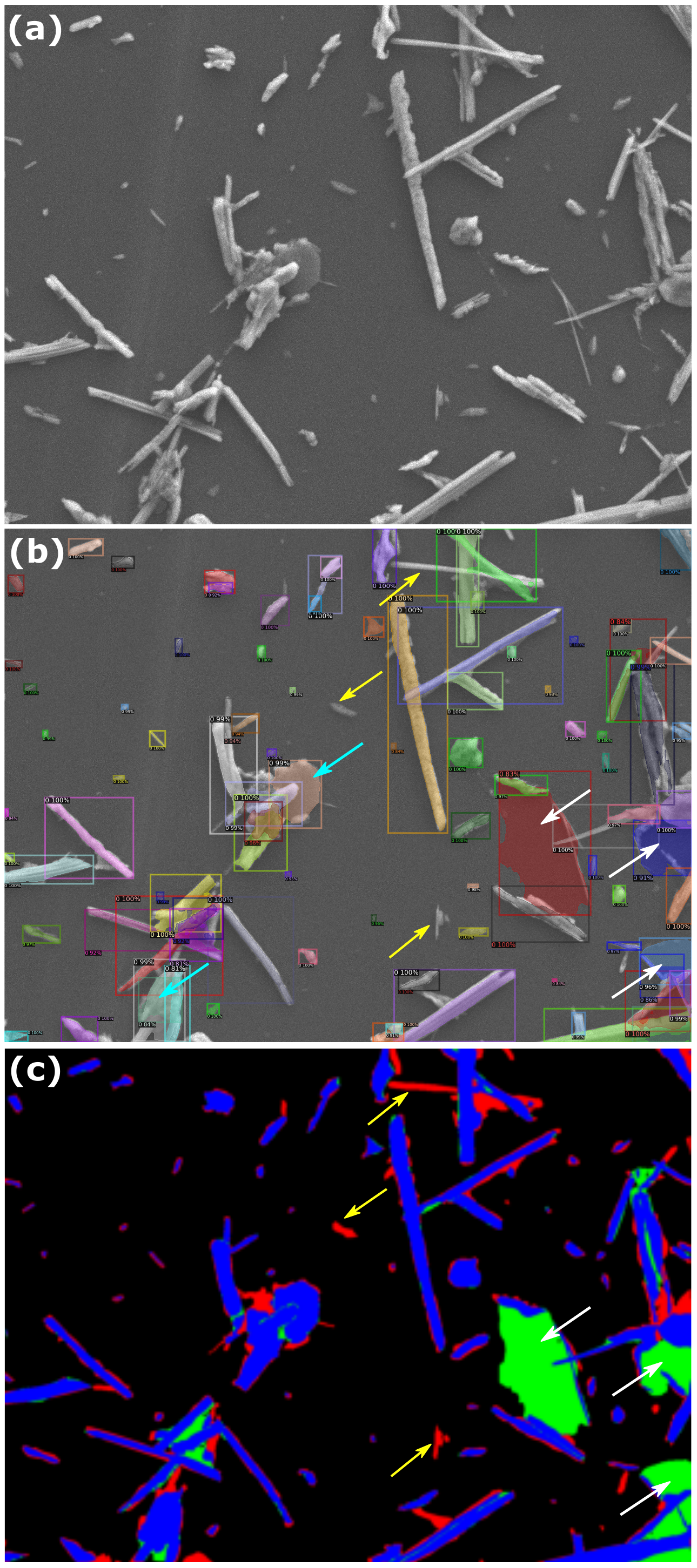}
  \caption{Model prediction on V$_2$O$_5$ nanowires within SEM image (957x1280 pixels). The best performer in AP segmentation mask in Tab.~\ref{tab:score_SEM} is used to visualize the prediction masks in (b).}
  \label{fig:Seg_SEM}
\end{figure}

Before applying the model to real microscopy images, 20 models were trained with various hyperparameters  to examine their influence on the synthetic images, see Tab.~\ref{tab:coco_training}.  Subsequently, all 20 trained models were applied to segment the three types of microscopy image. The best performer in mask segmentation AP for each image type was selected to visualize the segmentation masks in Fig.~\ref{fig:Seg_syntheticData},\ref{fig:Seg_SXTM},\ref{fig:Seg_SEM}(b). We start to evaluate the model segmentation accuracy at a semantic level to estimate to what degree the model can segment the actual nanowires from the background. The inherent strength of instance segmentation is that it includes the subordinate functionality of semantic segmentation, where the semantic results can be immediately extracted from the mask predictions. As shown in Fig.~\ref{fig:Seg_syntheticData},\ref{fig:Seg_SXTM},\ref{fig:Seg_SEM}~(b), fairly good segmentation results were obtained. These object masks were then used to evaluate semantic segmentation accuracy presented in Fig.~\ref{fig:Seg_syntheticData},\ref{fig:Seg_SXTM},\ref{fig:Seg_SEM}~(c). The color blue denotes the true positive pixels, to which the model predicts correctly as given in the ground truth provided by human annotation. The green color indicates false positive pixels, which means the model has inaccurately predicted that these pixels belong to a particular nanowire. The red color denotes the false negatives, which depict the pixels that belong to a nanowire (based on the ground truth) but were not identified as such by the model.  The performance of the trained model is discussed in more detail in the following sections.

Note that to make the model generally accessible as a segmentation tool of nanorod-like structures, we have further developed a web-based interactive application for readers to access and data-mining their own image datasets. Upon uploading the data and initializing the prediction model, statistics on predicted masks can be obtained and visualized accordingly. Details and access on the web-based interactive application can be found in data and code availability section.

\textbf{Synthetic nanowire image.} As for the synthetic nanowires shown in Fig.~\ref{fig:Seg_syntheticData}, not surprisingly, the deep learning model correctly segments particles contained within the synthetic nanowire datasets. The obtained AP of the tested 20 models reach a AP score of around 90 for bounding box regression and mask segmentation, respectively, see Tab. \ref{tab:coco_training}. As the synthetic image type is basically known to the model, the good accuracy are expected. Tuning the hyperparameter, dataset would generally result in a better AP in bounding box prediction and segmentation masks, however, not influence the results on synthetic image significantly.

\begin{table*}[ht]
  \caption{Model performance on X-ray pytchography image (Bounding box/Segmentation mask)}
  \label{tab:score_xray}
  \begin{tabular}{cccccccc}
    \toprule
    No. & Accuracy & AP  & AP$_{50}$ & AP$_{75}$ & AP$_{s}$ &  AP$_{m}$ & AP$_{l}$  \\
    \midrule
    13 & 86.6 &39.145/42.327 & 64.638/62.519 & 43.965/39.964 & 25.248/22.442 & 52.753/58.03 & 72.525/85.05 \\
    17 & 85.5 &41.809/41.719 & 74.603/67.042 & 43.25/40.438 & 28.34/19.785 & 52.772/52.774 & 62.624/70.099 \\
    14 & 86.2 &38.584/41.662 & 63.618/62.803 & 42.159/42.159 & 25.248/26.931 & 44.998/50.644 & 75.05/70.099  \\
    \bottomrule
  \end{tabular}
\end{table*}
ggests\begin{figure*}[h]
  \centering
  \includegraphics[width=0.9\textwidth]{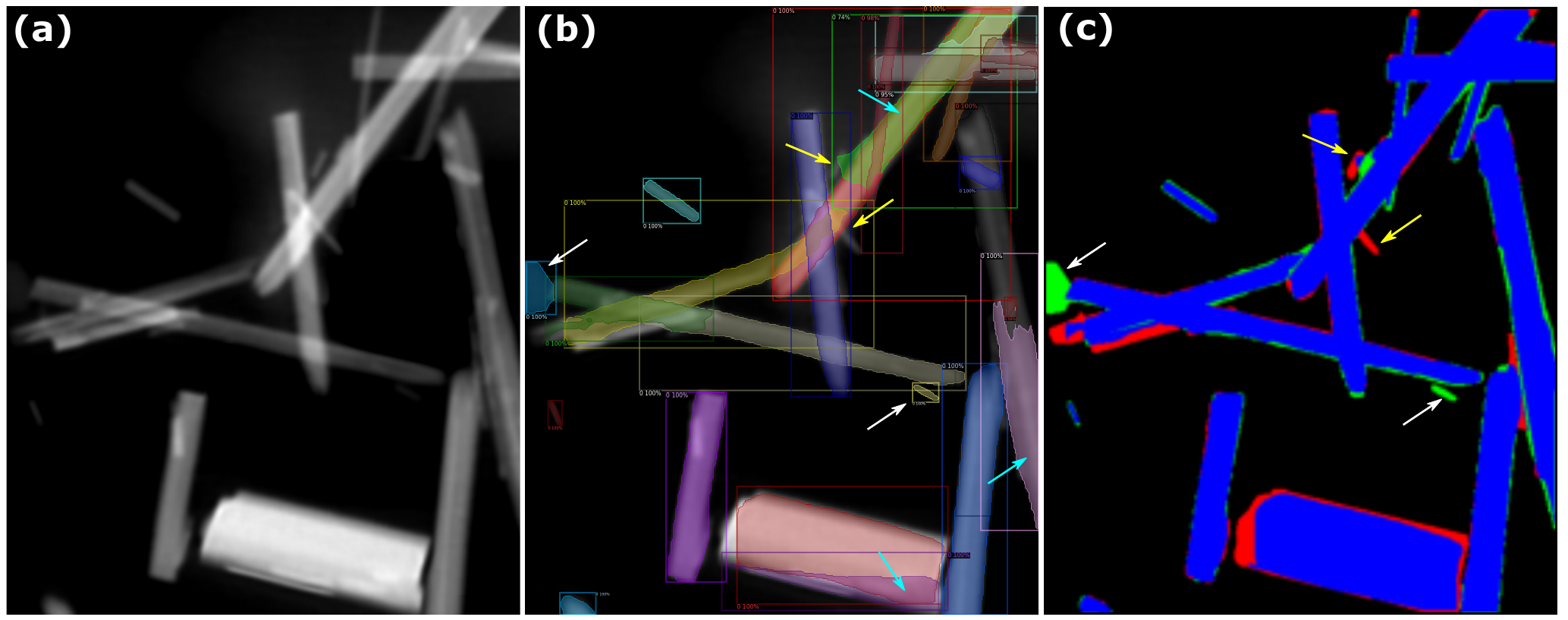}
  \caption{Model prediction on V$_2$O$_5$ nanowires within X-ray pytchography image (531x449 pixels). The best performer in AP segmentation mask in Tab.~\ref{tab:score_xray} is used to visualize the prediction masks in (b)}
 \label{fig:Seg_Xray}
\end{figure*}
\begin{figure*}[h]
     \centering
        \includegraphics[width=1\textwidth]{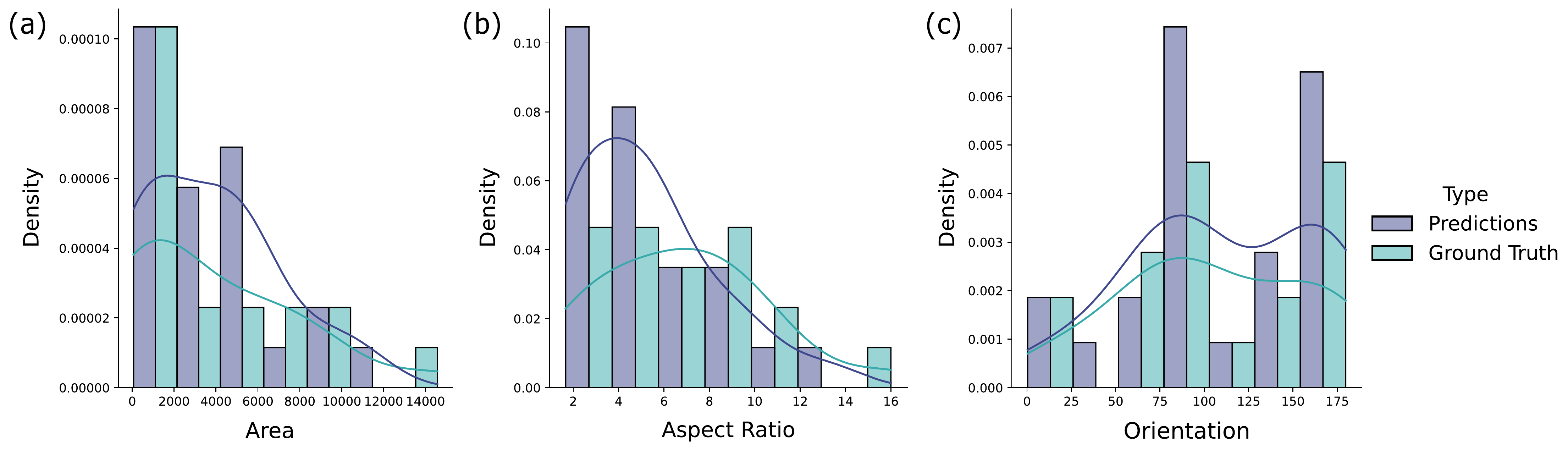}
        \caption{Particle statistics from the X-ray ptychography image are illustrated by the statistical density as a function of area (summation of pixels corresponding to each particle mask), aspect ratio (particle length/width), and orientation (angle relative to the horizontal axis) in (a), (b), and (c), respectively.}
        \label{fig:ST-Xray}

\end{figure*}

\textbf{X-ray ptychography image.} Although the model has been trained solely by synthetic datasets, good segmentation results are observed from experimental datasets. For X-ray ptychography images, the model predicts the overall binary mask with good accuracy and scores a segmentation score of 86.6, see Tab.~\ref{tab:score_xray}. The AP, AP$_{75}$ for both the bounding box, and the segmentation mask are comparatively high and score around 40, respectively. AP$_{50}$ reaches a score of around 62. From the metrics, it is mentionable that AP$_l$ are greater than AP$_s$ and AP$_m$, indicating that the image contains larger particles and they were segmented to a greater degree than smaller ones. At the instance level, two false-positive nanowires have been identified and are (shown in green) indicated by white arrows in Fig.~\ref{fig:Seg_Xray}(c); the origin of this false-positive result is low pixel intensity near the threshold that separates particles from background. For the same reason, these particles were not manually annotated but nevertheless were identified by the model thus demonstrating its performance, which is competitive with careful human annotation (while being much more accurate). The two particles shown in red (indicated by yellow arrows) are missed by the model, presumably by the noise in the corresponding region. But overall, considering the optical density input, the individual particles are extracted with good accuracy. In general, the mask predictions consistently perform extremely well when particles are well separated in space while overlapping regions (notoriously more difficult to segment) are still identified with good accuracy.

Some issues arise when the optical density gradient is low with less clear transitions in the overlapped region. Some issues arise when the optical density gradient is low with less clear transitions in the overlapped region. Further, the model tends to find smaller particles within larger instances as shown by the cyan arrow. The origin of this limitation stems from the broad range of particle aspect ratios and thicknesses in the experimental data (ca. 50 – 500nm), thus resulting in a less continuous optical density distribution with individual particles. In the training data, particle morphologies were generated assuming a prismatic structure with little-to-no variation in the cross-sectional shape. In contrast, the experimentally synthesized V$_2$O$_5$ are subject to defect formation, particle sintering, and intrinsic variations in the crystal growth during synthesis. This leads to a particle dispersion that is highly complex, non-prismatic and an ambiguous optical density mapping and further complicates the detection in the overlapped area and create artifacts which would possibly "fool"  trained models. This also introduces additional challenges during synthetic data generation and the segmentation tasks.  Nevertheless, this complexity in particle size, shape, and extent of curvature has pronounced effects on the emergent properties of these cathode particles so their correct identification remains important~\cite{santos2020bending,andrews2020curvature,horrocks2013finite}. Here, over-predicted particle masks inside the larger ones can be easily removed in a post-processing step. This step was not performed here to preserve the originality of the model prediction. However, to enhance the general prediction capability of the model and avoid post-processing procedures, the generation of non-prismatic structures for training datasets can be instrumental and will remain as future work.

Lastly, statistical information on particle area size, aspect ratio and orientation are compared in Fig.\ref{fig:ST-Xray} in form of histogram with kernel density estimation. The area size is determined by summation of the pixels belonging to different particle masks. The aspect ratio is calculated as the ratio between image coordinates of the longer edge to the shorter edge of the corresponding predicted mask. The orientation is considered as the angle between the particle alignment to the horizontal axis and ranging from 0 to 180 degree. As can be found in Fig.\ref{fig:ST-Xray} shows the statistical information from the X-ray ptychography data is in a  qualitatively good agreement. The main discrepancy is contributed by the number of additionally detected smaller particles inside the larger particles and as explained previously, leading to a higher density of the histogram for area size of 2000-4000 pixels and aspect ratio 2-4. This perturbation can be also observed for orientation for particles with 75 to 100 degree and 150 to 175 degrees. As the number of particles in the image is comparatively small, the feature distribution is becomes sensitive to the number of detected particles.

%%%%%%%%%%%%%%%%%%%%%%%%%%%%%%%%%%%%%%%%%%%%%%% STXM
\begin{table*}[ht]
  \caption{Model performance on STXM image (Bounding box/Segmentation mask)}
  \label{tab:score_stxm}
  \begin{tabular}{cccccccc}
    \toprule
     No. & Accuracy & AP & AP$_{50}$ & AP$_{75}$ & AP$_{s}$ &  AP$_{m}$ & AP$_{l}$  \\
    \midrule
    1 & 75.6 &  27.267/21.791 & 54.404/51.877 & 23.423/17.587 & 25.228/20.862 & 37.744/28.877 & NaN \\
    3 & 73.1 &30.51/21.36 & 53.577/52.172 & 29.092/15.618 & 24.447/17.329 & 48.116/32.782 & NaN  \\
    9 & 75.3 &28.293/21.049 & 51.763/48.671 & 27.977/17.437 & 20.581/16.013 & 49.245/34.714 & NaN \\
    \bottomrule
  \end{tabular}
\end{table*}
\begin{figure*}[h]
     \centering
         \includegraphics[width=1.0\textwidth]{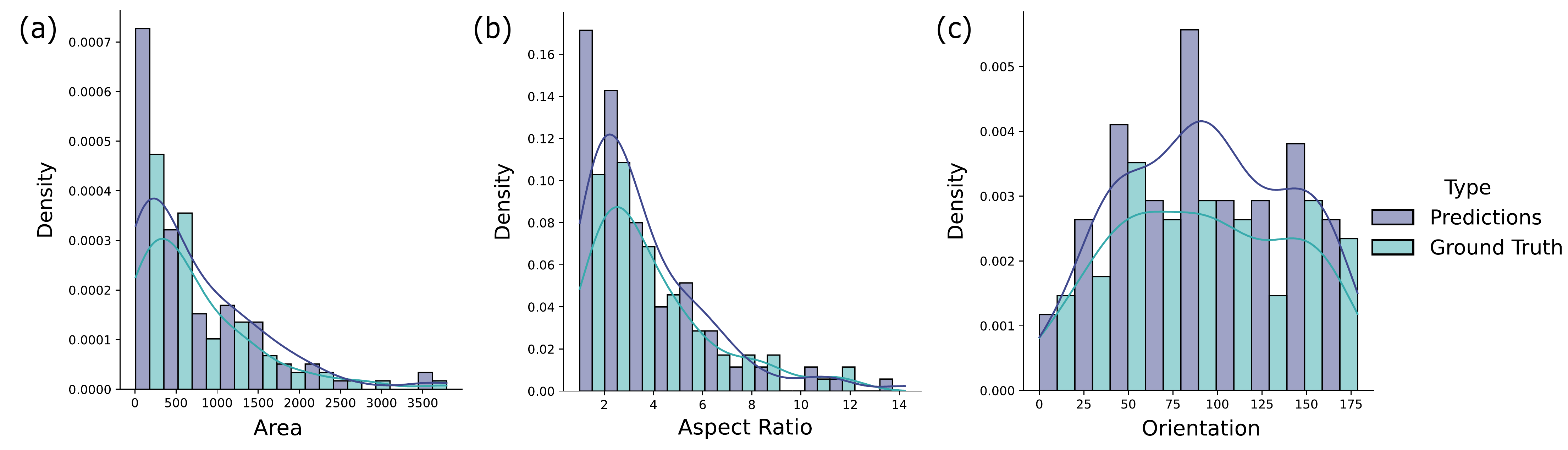}
        \caption{Statistical distribution of features in the STXM image}
        \label{fig:ST-STXM}
\end{figure*}

\textbf{STXM image.} Of the two X-ray microscopy techniques considered in this work, X-ray ptychography offers the greatest spatial resolution (ca. 6 nm), thus, from a purely image segmentation perspective we expect the performance of the model to be greatest for this class of images. Nevertheless, techniques such as scanning transmission X-ray microscopy which offer slightly lower spatial resolution (ca. 25 nm) but enabled more detailed mapping of spectral features (i.e. have richer chemical information) are equally important for cheminformatics. In this work, the original resolution of the STXM image was 100x100 pixels and fewer than in the X-ray pytchography image. To enable sharper manual annotation, we rescaled the STXM image to the size of pytchography image for easier visual access (it is important to note that this does not fundamentally change the resolution enabled by the experimentation). The number of particles, their variations in morphology, and the complexity of their dispersion is noticeably greater than the X-ray ptychography image making the segmentation task considerably more challenging. Nevertheless, the segmentation accuracy scores around 75 and the AP score is ca. 21. Since there exist relatively no larger particles in the image, AP$_l$ was not provided. The comparably lower but surprisingly good scores highlight the complexity of segmenting complex particle dispersions with several instances over overlap and agglomeration This is especially noticeable for the false positive green particles in
 Fig.~\ref{fig:Seg_SXTM}(c), w) which were generally overlooked in the manual annotation process. The access of human annotation is strictly limited for image of such complexity. However, at the instance level, from a visual perspective, the model performs considerably well. Overlapped particles are consistently identified and agglomerations, while difficult to identify visually, are captured by the deep learning model. The statistical distribution of the features in Fig.~\ref{fig:ST-STXM} agree well both to a qualitative and quantitative extent. The shape of kernel density estimation agrees well with the manual ground truth distribution. As the model prediction captured smaller particles, which were not manually labelled (further demonstrating the performance of the model), the statistics of the prediction are shown to have a higher density distribution in each feature characteristic, observable as small peak shift of the KDE curve. The results suggest that for complicated particle networks contained in a relatively low-resolutioned and low-contrast image, deep learning models indeed deliver more comprehensive information on the statistical information than human annotations.

%%%%%%%%%%%%%%%%%%%%%%%%%%%%%%%%%%%%%%%%%%%%SEM
\begin{table*}[ht]
  \caption{Model performance on SEM image (Bounding box/Segmentation mask)}
  \label{tab:score_SEM}
  \begin{tabular}{ccccccccc}
    \toprule
     No. & Accuracy & AP & AP$_{50}$ & AP$_{75}$ & AP$_{s}$ &  AP$_{m}$ & AP$_{l}$  \\
    \midrule
    1 & 59.6 & 27.61/12.927 & 51.8/23.126 & 22.884/5.206 & 20.586/7.842 & 43.495/22.443 & 38.356/36.634 \\
    5 & 61.7 &27.916/12.688 & 54.302/37.916 & 21.149/5.643 & 22.776/8.337 & 43.195/21.884 & 16.311/25.743 \\
    13 & 33.2 &21.561/12.659 & 47.924/35.553 & 15.718/5.088 & 14.011/8.089 & 44.587/23.695 & 15.317/12.871 \\
    \bottomrule
  \end{tabular}
\end{table*}

\begin{figure*}[h]
     \centering
         \includegraphics[width=1.0\textwidth]{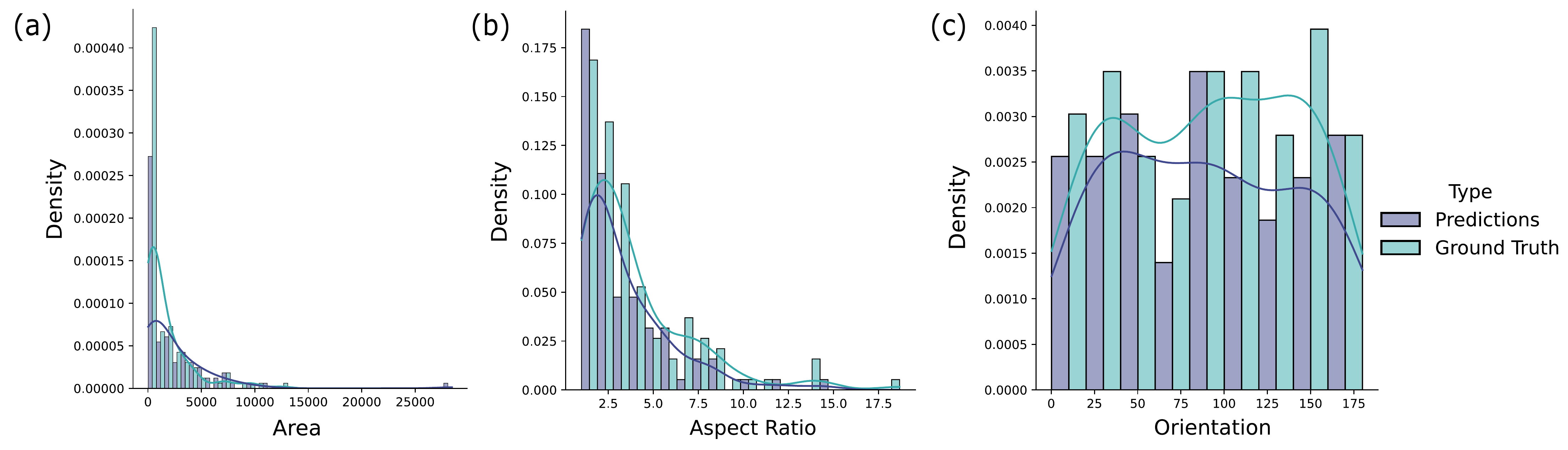}
        \caption{Statistical distribution of features in the SEM image}
        \label{fig:ST-SEM}
\end{figure*}

\textbf{SEM image.} The deep learning model shows success in segmenting the particles shown in the ptychography and STXM optical density images, in part, due to the mechanisms of contrast generation, which involves the transmission of an incident X-ray source through the bulk of the material. Here, the degree of transmission is related to the energy-specific elemental absorption cross-section, corresponding to excitation of electrons from core levels to unoccupied or partially occupied states, giving rise to similar absorption contrast for compositionally homogeneous particles and allowing the discernment of overlapped intersections (in the form of increased optical density due to thickness effects). As a point of comparison, in scanning electron microscopy, the detection of secondary electrons or backscattered electrons from the surface is sensitive to surface morphology, edge effects, and charge build-up and is fundamentally different from the X-ray ptychography and STXM images shown in previous sections. To demonstrate the versatility of the model, we introduce a scanning electron micrograph for the purposed of segmentation. Despite the fundamental differences in contrast generation between the data used to train the model and the SEM data utilized as an input here, the model performs surprisingly well and is still able to identify overlapping regions despite the absence of optical density information. Here, the deep learning model captures the contrast gradients at the particle boundaries and utilizes it as a criterion to identify individual fibers in the overlapped regions, independent of background. Much like the previous datasets, agglomerations cannot be well separated as shown by the cyan arrow in Fig.~\ref{fig:Seg_SEM}(b). In addition to unidentified nanowires (indicated by  yellow arrows), some issue exists in the right region of the image, where larger masks (green, false-positive pixels, indicated by white arrows) were predicted for individual fibers, presumably due to the perspective of the taken image, causing different contrast in the image. To improve the observed inconsistency, an additional class for agglomerated phase can be introduced and generated in the microstructure generation step, to further differentiate between different foreground phases and, subsequently, nanowires instances. Despite the lower AP score around 13, the statistical results seem to be less sensitive in the presence of more significant particle numbers. The main deviation lies in the number of undetected isolated particles of
smaller size and therefore leading to an underestimation of statistical density w.r.t area size up to 1000 pixels, aspect ratio from 2.5 to 7.5, and orientation from 100 - 150 degrees, as shown in Fig.~\ref{fig:ST-SEM} (a-c).

\section{Conclusion and outlook}

Advancements in chemical imaging have enabled improvements in image collection speed, signal-to-noise, and computational power enabling the stronger connections between structure-function relationships and emergent properties. In this work, we have developed a deep learning model for instance segmentation which has been trained solely by synthetic datasets emulated after experimental microscopy data. A benchmarked assessment of the Mask R-CNN algorithm assesses the prediction accuracy of the model on experimental datasets collected by X-ray ptychography, scanning transmission X-ray microscopy, and scanning electron microscopy. Despite variations in spatial resolution, particle dispersion densities, and contrast generation, the model scores well across all the considered microscopy data thus demonstrating the versatility of the neural network in the segmentation tasks. The introduction of optical density values in the synthetic datasets utilized to train the model enables accurate predictions and individual segmentation instances for overlapping particles - a challenge which is strongly relevant to particle dispersions but has seen limited advancement due to inherent complexities. A web-based, and interactive segmentation tool based on the model developed in this work has been made publicly available at \url{https://share.streamlit.io/linbinbin92/V2O5_app/V2O5_app.py}. Future work will focus on the introduction of noise to replicate light perturbations, angle dependencies, and variable background characteristics in order to improve the models robustness to real-life datasets. We will further seek to implement the developed methods in real-time process control settings during nanowire growth and battery operation.

\section{Data and code availability}
The code and dataset developed in this work can be found in \url{https://github.com/linbinbin92/V2O5_app/tree/master}. The interactive web-based segmentation application can be found: \url{https://share.streamlit.io/linbinbin92/V2O5_app/V2O5_app.py}

\section{Acknowledgments}
This work was supported by German Research Foundation (DFG) B. Lin, B-X. Xu acknowledge the financial support under the grant agreement No. 405422877 of the Paper Research project (FiPRe) and National high performance computing center for computational sci- ences(NHR4CES). We acknowledge Abhishek Parija and Dr. David Schapiro for their assistance with X-ray ptychography experiments. The authors also greatly appreciate their access to the Lichtenberg High Performance Computer and the technical supports from the HHLR, Technical university Darmstadt. X-ray ptychography measurements were performed at the COherent Scattering and MICroscopy (COSMIC) branch of the Advanced Light Source (ALS). A portion of the STXM measurements utilized in this work was collected at the Canadian Light Source, which is supported by the Natural Sciences and Engineering Research Council of Canada, the National Research Council Canada, the Canadian Institutes of Health Research, the Province of Saskatchewan, Western Economic Diversification Canada, and the University of Saskatchewan. The research at Texas A\&M University was supported by the NSF under DMR 1627197. D.A.S. acknowledges support under a NSF Graduate Research Fellowship under grant No. 1746932.

\section{Author contribution}
B.L. conceived the research. B.L and N.E performed the dataset generation and model training and the formal analysis. D.A.S was involved in the conceptualization of the project. Y.L prepared the particles shown in the images. B.L, N.E, D.A.S drafted the manuscript. B.X.X and S.B reviewed and discussed the results and organized the funding.

\bibliographystyle{ieeetr}
\bibliography{sample}

\end{document}